\documentstyle[preprint,aps]{revtex}
\tighten
\begin{document}
\draft
\title{Dynamic structure of He-Ne mixtures by MD simulation: from hydrodynamic to fast and slow sound modes}
\author{Marco Sampoli,$^{1,2}$ Ubaldo Bafile,$^{2,3}$ Eleonora Guarini,$^2$ and 
Fabrizio Barocchi,$^{2,4}$}
\address{$^1$ Dipartimento di Energetica "S. Stecco", Universit\`{a} di Firenze, via di S. Marta 3, I-50139 Firenze, Italy \\
$^2$ Istituto Nazionale per la Fisica della Materia, Unit\`{a} di Ricerca di Firenze, largo E. Fermi 2, I-50125 Firenze, Italy\\
$^3$ Istituto di Elettronica Quantistica, Consiglio Nazionale delle Ricerche, via Panciatichi 56/30, I-50127 Firenze, Italy \\
$^4$ Dipartimento di Fisica, Universit\`{a} di Firenze, largo E. Fermi 2, I-50125 Firenze, Italy\\}

\date{\today}

\maketitle

\begin{abstract}

Molecular dynamics results for the dynamic structure of a He$_{0.77}$Ne$_{0.23}$ gas mixture at two densities (15.8 nm$^{-3}$ and 36.1 nm$^{-3}$) are presented. A clear description of the crossover from hydrodynamic modes to distinct excitations for the two species is obtained. The higher density data neatly show high- and low-frequency branches setting on in the dispersion curve, with a rather localized transition. The lower density results remarkably agree with existing neutron scattering data and, differently from previous simulation studies, display hydrodynamic behavior up to $k \approx 2$ nm$^{-1}$. A smooth transition to fast sound is shown to take place for $2<k$/nm$^{-1}<5$, where the present MD data fill the existing gap in the experimental results.
\end{abstract}  


\narrowtext
\section*{}
Inelastic neutron scattering studies of the dynamic behavior of two-component fluids were usefully performed on dense He-Ne and He-Ar gas mixtures \cite{1,2,3,4,5}. Both conventional \cite{1,2,3} and, recently, Brillouin \cite{5} neutron spectroscopy produced, in particular, a rather extended set of data for the dynamic structure factor $S(k,\omega)$ of He-Ne mixtures with low concentration (20-30\%, typically) of the heavier component, and total number density $n$ around 16 nm$^{-3}$. Important and detailed indications on the nature of dynamic excitations in these mixtures have been obtained from the above neutron data. On the one hand, they show that at wavevectors $k>4$ nm$^{-1}$ \cite{1} the total $S(k,\omega)$ contains two Brillouin doublets, one at frequencies $\pm\:\omega_{\sc s}^{(1)}$, corresponding to a "fast" sound propagating with velocity $c_1$ nearly equal to that of pure He at the same temperature and total density of the mixture ($\sim$ 540 m s$^{-1}$), and one, not giving rise to visible inelastic features in $S(k,\omega)$, at lower frequencies, $\pm\:\omega_{\sc s}^{(2)}$, attributed to the presence of a "slow" sound related to the Ne atoms dynamics (as in previous papers, labels 1 and 2 refer to He and Ne, respectively). On the other hand, from high resolution neutron Brillouin spectra at $k<1.8$ nm$^{-1}$ \cite{5} no decoupling of the dynamics of the two species is observed. Instead, one single acoustic mode was detected, with a linear dispersion in remarkable agreement with the hydrodynamic sound velocity $c \approx 370$ m s$^{-1}$ of the mixture, as estimated from the van der Waals equation \cite{3}. There is, therefore, experimental evidence that the transition from hydrodynamic behavior towards distinct excitations for the two components occurs in the wavevector region around $k\sim 3$ nm$^{-1}$, for this He-Ne state. 

This experimental result, however, does not agree with the analysis of a molecular dynamics (MD) simulation of the same mixture \cite{6}, where the onset of a lower frequency mode in the Ne dynamics was detected at a $k$-value as small as 0.6 nm$^{-1}$ by analyzing the partial correlations. At larger $k$ values, the MD data of Ref. \cite{6} indicate a progressive departure from hydrodynamic sound (more evident in the Ne case), characterized by split, low- and high-frequency, branches for the heavy and light component, respectively. Very recently, new simulation results on the same mixture were also announced \cite{7} which, while agreeing with the experimental data of Ref. \cite{5} for the total $S(k,\omega)$, again confirm the findings of the older simulation.

In order to investigate this apparent discrepancy between neutron Brillouin and MD results, we performed extended simulations of two He$_{0.77}$Ne$_{0.23}$ mixtures: one in the same thermodynamic state ($n=15.8$ nm$^{-3}$, $T=39$ K) as in the quoted neutron \cite{1,5} and MD \cite{6,7} studies; the other in a far denser state ($n=36.1$ nm$^{-3}$, $T=60$ K), because, with increasing density, hydrodynamics is expected to hold in a wider $k$ range, and the features of the Rayleigh-Brillouin spectrum should be enhanced by the reduced line widths and by the higher speed of sound in the mixture, both effects allowing for a clearer detection of dynamic decoupling of the two components.

The constant-$NVE$ simulations were carried out by using 87808 particles confined in a cubic box of 17.7 nm side at the lower density (13.4 nm side at the higher density) with periodic boundary conditions. The equations of motion were integrated using the leap-frog algorithm with a time step of 20 fs at low density and of 12.5 fs at high density. After 10$^4$ time steps for thermalization, the dynamics has been followed for about 2400 ps and 600 ps, at low and high density, respectively. The time step for configuration recording was fixed to 0.1 ps. The corresponding wavevector ranges were $0.4<k$/nm$^{-1}<5$, at low density, and $0.5<k$/nm$^{-1}<6.6$, at high density. The He-He and Ne-Ne interactions were modeled according to the Aziz et al. potentials \cite{8,9}, while the He-Ne cross interaction parameters were derived by applying the Lorentz-Berthelot rules \cite{10}. 

Here we first summarize the high density simulation results. At low $k$, i.e. in the hydrodynamic regime, each $S_{ij}(k,\omega)$ ($i$, $j$ = 1, 2) is predicted to be composed of a central line, resulting from the superposition of two Lorentzian components (both located at zero frequency) related to concentration diffusion and heat diffusion, and of two Lorentzian lines, with asymmetric correction, centred at $\pm\:\omega_{\sc s}$, i.e. the typical Brillouin doublet \cite{3,10}. In formulae:

\begin{equation} 
S(k,\omega)=\frac{1}{\pi}\left[\frac{a_{01}z_{01}}{z_{01}^{2}+\omega^2}+\frac{a_{02}z_{02}}{z_{02}^{2}+\omega^2}+\frac{a_{\sc s}z_{\sc s}+b_{\sc s}(\omega+\omega_{\sc s})}{z_{\sc s}^{2}+(\omega+\omega_{\sc s})^2}+\frac{a_{\sc s}z_{\sc s}-b_{\sc s}(\omega-\omega_{\sc s})}{z_{\sc s}^{2}+(\omega-\omega_{\sc s})^2}\right]
\end{equation}

Other models based on extended sets of dynamical variables have been applied in theoretical studies of He-Ne mixtures \cite{2,11}. However, the eight-parameter fit of Eq. (1), convoluted with the simulation resolution function, to the MD data provides an excellent description of all $S_{ij}(k,\omega)$ in the whole $k$-range of our simulations. In the upper frames of Fig. 1, $S_{11}(k,\omega)$ and $S_{22}(k,\omega)$ are plotted at three $k$ values, together with the fitted lines. Logarithmic scales are used to enhance the visibility of the inelastic features. The frequency range in the plots is limited to three times the fitted  $\omega_{\sc s}^{(1)}$, though the used fit range is larger. We note that Eq. (1) proves very good over more than 4 orders of magnitude in intensity. The lower frames show both data and fitted curves multiplied by $\omega^2$, i.e. the longitudinal current spectra $C_{ij}(k,\omega)=\omega^2 S_{ij}(k,\omega)$, including the cross term $C_{12}(k,\omega)$. The following features can be observed. The Brillouin peak in $S_{11}(k,\omega)$ is visible at all $k$, while that of $S_{22}(k,\omega)$ appears only as a very weak shoulder at the intermediate $k$, and can no longer be seen at the highest one. $C_{ij}(k,\omega)$ presents another structure at a lower frequency, of the order of the width of the central line of $S_{ij}(k,\omega)$, i.e. where interdiffusion processes dominate the spectrum. The main feature of $C_{11}(k,\omega)$ is, at all $k$, the Brillouin peak, while in the Ne-Ne correlation the intensity evolution of the two peaks as a function of $k$ determines the position of the maximum of $C_{22}(k,\omega)$: at low $k$, this is close to $\omega_{\sc s}^{(2)}$, but at higher $k$, where the Brillouin intensity becomes too low to produce a maximum, it shifts to the location of the lower frequency peak. Inspection of the cross correlation reveals that, in the first case, the maximum of $C_{22}(k,\omega)$ occurs where $C_{12}(k,\omega)$ is positive, indicating in-phase motions of the two components, while, in the second case, it is located in a frequency range that cannot be related to an acoustic propagation, since it is characterized by negative values of $C_{12}(k,\omega)$, a signature of inter-species diffusion dynamics.
 
The dispersion curves $\omega_{\sc s}(k)$ of the fitted inelastic peak frequency in $S_{11}(k,\omega)$ and $S_{22}(k,\omega)$ are shown in Fig. 2. Up to $k \approx 4$ nm$^{-1}$, the He and Ne curves clearly show the same (substantially linear) behavior with a slight positive dispersion: the collective excitations propagate with speed $c_1=c_2= c$ varying from 900 to 1000 m s$^{-1}$. Above this $k$ value, the dispersion curve clearly splits into two branches, indicating the onset of a dynamic decoupling of the two species that gives rise to fast and slow sound in the mixture, with $c_1=\omega_{\sc s}^{(1)}/k$ rising up to 1050 m s$^{-1}$ and $c_2=\omega_{\sc s}^{(2)}/k$ decreasing sharply down to 800 m s$^{-1}$, at $k=6.5$ nm$^{-1}$. These values are in reasonable agreement with the speed of sound of the pure components (at the same temperature and total density of the mixture), i.e. 1150 m s$^{-1}$ for He \cite{12} and 840 m s$^{-1}$ for Ne \cite{13}.

Fig. 2 also shows the dispersion curve derived as the position of the maximum in $C_{ij}(k,\omega)$, an often used procedure \cite{6,7}. In the He case such a determination provides a frequency larger than $\omega_{\sc s}^{(1)}$, while for the Ne-Ne correlation this method produces, for the reason explained before, a wrong dispersion curve which departs from linearity at rather small $k$ values, with frequencies much lower than $\omega_{\sc s}^{(2)}$, and in the characteristic range of the concentration diffusion processes. The wiggles in such a curve, evident in Fig. 2, correspond to a $k$ range where the two peaks in $C_{22}(k,\omega)$ have comparable heights, so that its maximum jumps from one position to the other due to statistical noise. 

We now turn to the low density simulation results, for which we repeated the partial structure factors analysis described above, obtaining, on the whole, a similar picture. Again, Eq. (1) remarkably accounts for all $S_{ij}(k,\omega)$ at all $k$ values, as shown in Fig. 3. The inelastic features of each $C_{ij}(k,\omega)$, including the sign of the cross term, display a $k$ dependence similar to that of the high density mixture. However, in the Ne case the Brillouin peak amplitude vanishes for $k>2$ nm$^{-1}$ and prevents from a reliable determination of its position. Thus, the Ne-Ne inelastic structures cannot be followed further.

The fitted $\omega_{\sc s}^{(1)}$ and $\omega_{\sc s}^{(2)}$ as functions of $k$ are given in Fig. 4. No departure from hydrodynamic sound is observed for either component ($c_1=c_2=c \approx 370$ m s$^{-1}$) for all $k$ values below 1.9 nm$^{-1}$. For $2<k$/nm$^{-1}<5$, the propagation speed $c_1$ of the light particles excitations neatly undergoes a smooth transition from the hydrodynamic sound speed $c$ of the mixture to the quoted fast sound velocity ($\sim 540$ m s$^{-1}$). Since the total $S(k,\omega)$ is dominated by the $S_{11}(k,\omega)$ term and shows therefore inelastic peaks at frequencies similar to those of the He-He correlation, this result provides the missing link between the low- and high-$k$ neutron data (see Fig. 2 of Ref. \cite{5} and Fig. 3 of ref. \cite{1}).  

The dispersion curves obtained from the maxima of the longitudinal current spectra, also shown in Fig. 4, again give a different indication: they both would suggest a departure from the hydrodynamic straight line at $k$ around 0.7 nm$^{-1}$, with a deviation reaching 15\% around $k \approx 2$ nm$^{-1}$ for He \cite{14} and much larger (negative) values for Ne. Indeed, for the Ne case, such a curve is in good agreement with the analogous one in Fig. 4 of Ref. \cite{6}.

The molecular-dynamics investigation reported in this Letter clearly illustrates the onset of decoupled modes in dense gaseous binary mixtures, from which the fast sound phenomenon originates. The He-Ne mixtures here considered provide valuable insight on these effects. In the high density case, the splitting of the dispersion curve into two branches is a sharp effect and a slow sound propagation is clearly detected. At the lower density, a very good agreement is found with the existing neutron measurements. The present MD data smoothly bridge the gap between the low-$k$ neutron Brillouin results \cite{5} and the older high-$k$ experiment \cite{1}. In particular, hydrodynamic behavior up to $k \approx 2$ nm$^{-1}$ is confirmed, and a continuous transition to fast sound is shown to take place for $2<k$/nm$^{-1}<5$. From the present data it cannot be decided whether a lower branch of the dispersion curve also exists at the lower density, similarly to what happens at the higher one. It may be that the Ne dispersion curve follows that of He up to a $k$ value larger than 2 nm$^{-1}$, before the splitting takes place, or that no acoustic mode propagates through the Ne atoms above $k \approx 2$ nm$^{-1}$, as suggested by the vanishing amplitude of the Brillouin peak. Finally, it is clear from this work that, at least in the presence of very weak inelastic structures compared with quasi-elastic diffusion processes, an analysis based on the location of maxima in the longitudinal current spectra can lead to quite incorrect results.

\begin{figure}
\caption{MD spectra at the higher density and three values of $k$. Upper frames: $S_{11}(k,\omega)$ (dots) and $S_{22}(k,\omega)$ multiplied by 3 (circles). Lower frames: $C_{11}(k,\omega)$ (dots), $C_{22}(k,\omega)$ multiplied by 3 (circles), and $C_{12}(k,\omega)$ (squares). In all frames the solid lines are best fits with Eq. (1). Only few data points are shown for the sake of clarity. Error bars, where not visible, are smaller than the size of the symbols. The arrows in the lower frames show the values of $\omega_{\sc s}^{(2)}$.}
\end{figure}

\begin{figure}
\caption{Dispersion curves of the acoustic modes in MD $S_{11}(k,\omega)$ and $S_{22}(k,\omega)$ at the higher density. The fitted $\omega_{\sc s}^{(1)}$ (dots) and $\omega_{\sc s}^{(2)}$ (circles) are shown with error bars. Solid lines are obtained as positions of maxima of $C_{11}(k,\omega)$ (upper) and $C_{22}(k,\omega)$ (lower).}
\end{figure}

\begin{figure}
\caption{MD spectra at the lower density and three values of $k$. The format of the picture is the same as in Fig. 1. The arrow in the bottom right frame is missing since $\omega_{\sc s}^{(2)}$ cannot be obtained reliably from the fit for $k>2$ nm$^{-1}$.}
\end{figure}

\begin{figure}
\caption{Dispersion curves of the acoustic modes in MD $S_{11}(k,\omega)$ (top) and $S_{22}(k,\omega)$ (bottom) at the lower density. The fitted $\omega_{\sc s}^{(1)}$ (dots) and $\omega_{\sc s}^{(2)}$ (circles) are shown with error bars. Solid lines are obtained as positions of maxima of $C_{11}(k,\omega)$ (top) and $C_{22}(k,\omega)$ (bottom). Dashed straight lines correspond to propagation speeds of 374 and 544 m s$^{-1}$.}
\end{figure}

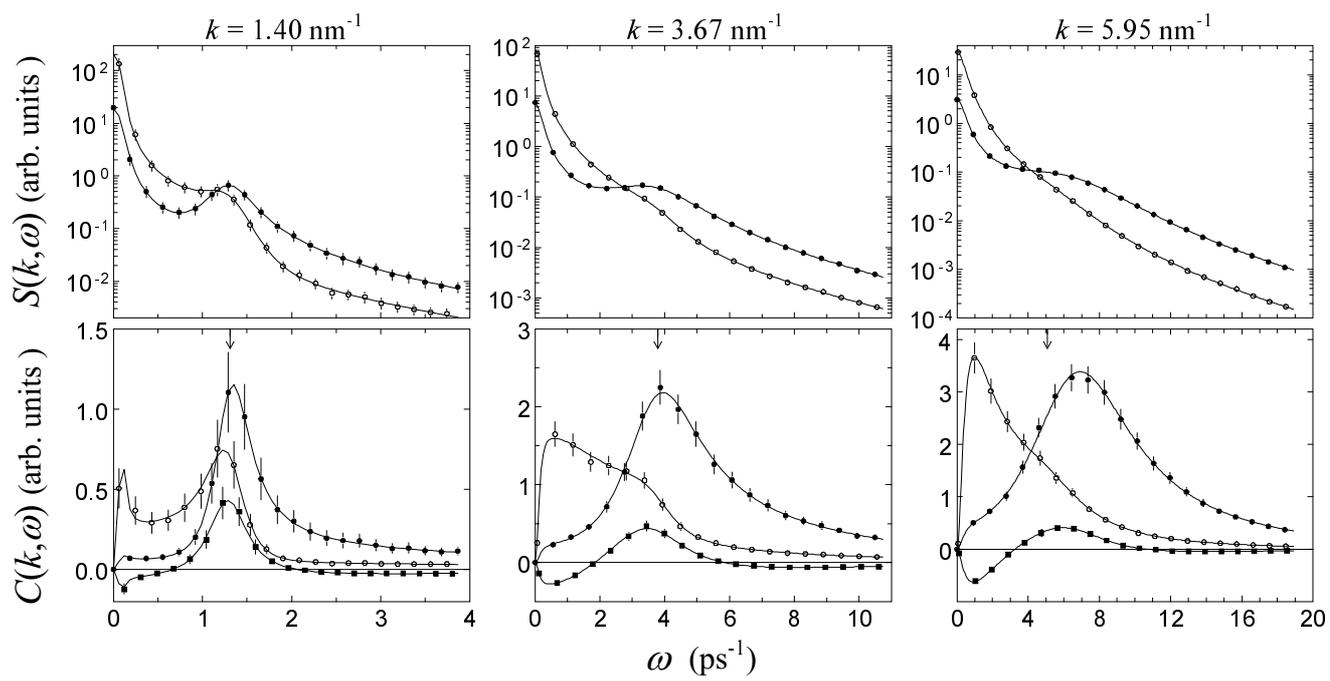
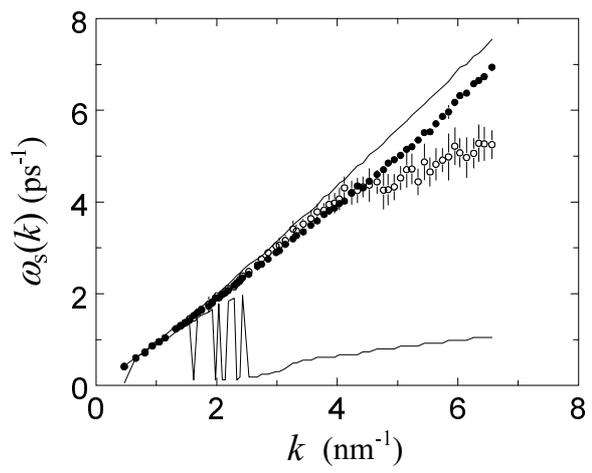
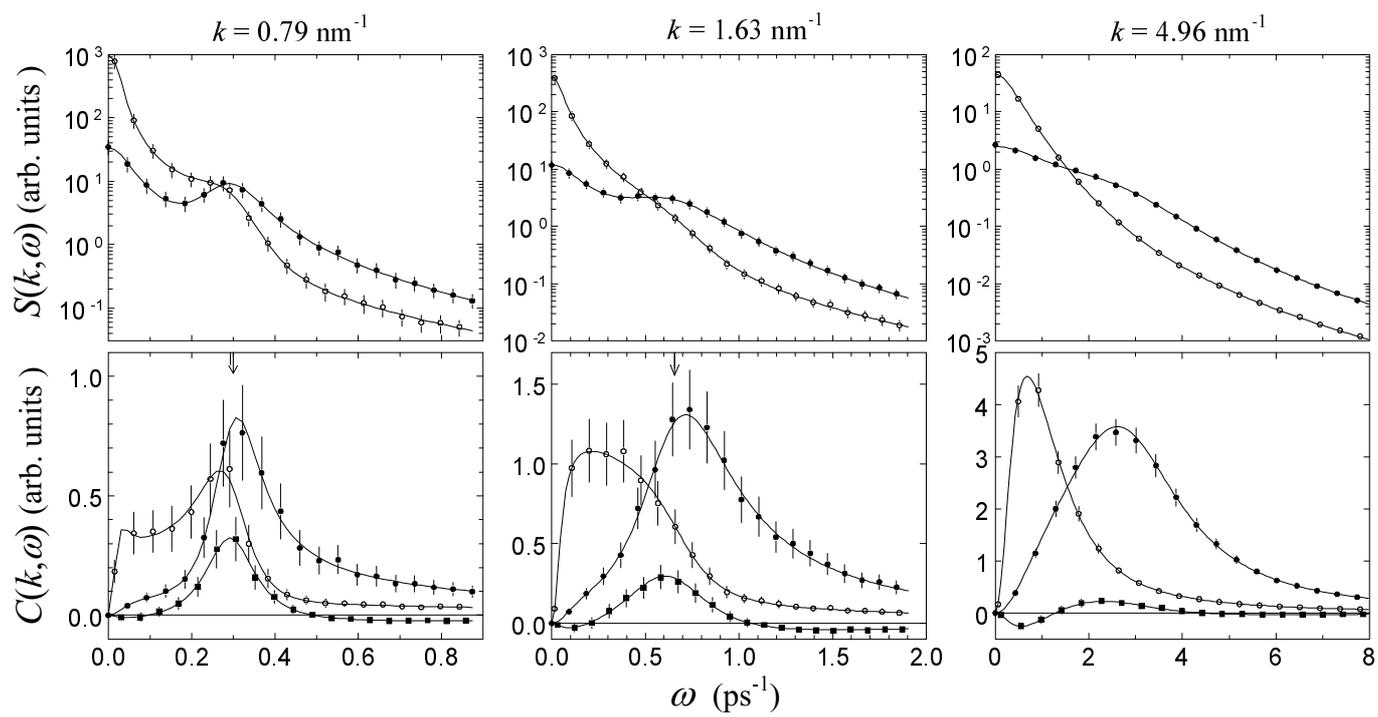
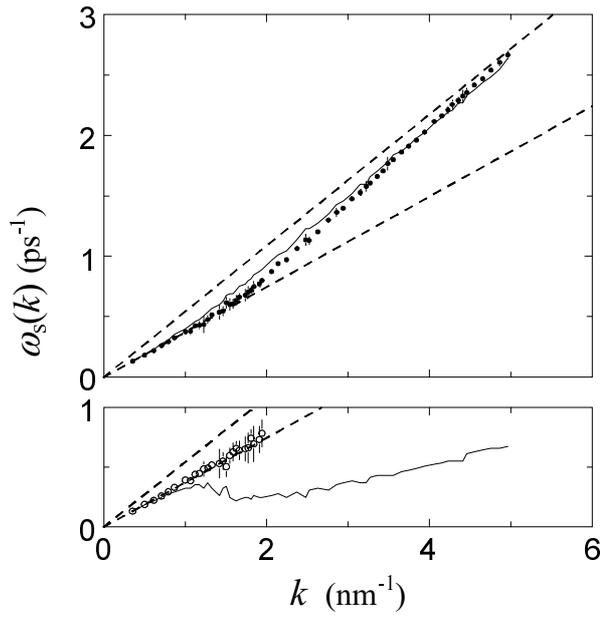
\end{document}